\def\BLIND{0} 
\def\ARXIV{1} 
\def\PAGENUMS{0} 
\def\AAM{0} 

\ifnum\ARXIV=1
    \def\PAGENUMS{1} 
    \def\AAM{1} 
\fi

\documentclass[a4paper,twoside]{article}

\usepackage{epsfig}
\usepackage{subcaption}
\usepackage{calc}
\usepackage{amssymb}
\usepackage{amstext}
\usepackage{amsmath}
\usepackage{amsthm}
\usepackage{multicol}
\usepackage{pslatex}
\usepackage{apalike}
\usepackage{algorithm2e}
\usepackage[bottom]{footmisc}

\usepackage{svg}
\usepackage[hidelinks]{hyperref}
\usepackage{url}
\usepackage{balance}
\let\bibhang\relax
\usepackage{natbib}
\usepackage{booktabs}
\usepackage{fancyhdr} 
\usepackage{lastpage} 

\usepackage{SCITEPRESS} 

\begin{document}

\title{BondBERT: What we learn when assigning sentiment in the bond market}

\ifnum\BLIND=1
    \author{Anonymous submission 
    \\ } 
\else
    \author{\authorname{Toby Barter\sup{1}\orcidAuthor{0009-0000-3888-8672}, Zheng Gao\sup{1}, Eva Christodoulaki\sup{1}\orcidAuthor{0000-0003-0099-6111}, Jing Chen\sup{2}\orcidAuthor{0000-0001-7135-2116} and John Cartlidge\sup{1}\orcidAuthor{0000-0002-3143-6355}}
    \affiliation{\sup{1}School of Engineering Mathematics and Technology, University of Bristol, Bristol, UK}
    \affiliation{\sup{2}School of Mathematics,
      Cardiff University, Cardiff, UK}
    \email{\{wu22796, ww24988, Eva.Christodoulaki\}@bristol.ac.uk, ChenJ60@cardiff.ac.uk, John.Cartlidge@bristol.ac.uk}
    }
\fi


\abstract{Bond markets respond differently to macroeconomic news compared to equity markets, yet most sentiment models are trained primarily on general financial or equity news data. However, bond prices often move in the opposite direction to economic optimism, making general or equity-based sentiment tools potentially misleading. We introduce BondBERT, a transformer-based language model fine-tuned on bond-specific news. BondBERT can act as the perception and reasoning component of a financial decision-support agent, providing sentiment signals that integrate with forecasting models. We propose a generalisable framework for adapting transformers to low-volatility, domain-inverse sentiment tasks by compiling and cleaning 30,000 UK bond market articles (2018--2025). BondBERT's sentiment predictions are compared against FinBERT, FinGPT, and Instruct-FinGPT using event-based correlation, up/down accuracy analyses, and LSTM forecasting across ten UK sovereign bonds. We find that BondBERT consistently produces positive correlations with bond returns, and achieves higher alignment and forecasting accuracy than the three baseline models. These results demonstrate that domain-specific sentiment adaptation better captures fixed income dynamics, bridging a gap between NLP advances and bond market analytics.} 

\onecolumn \maketitle \normalsize \setcounter{footnote}{0} \vfill

\ifnum\PAGENUMS=1
    \thispagestyle{fancy}
    \pagestyle{fancy}
    \fancyfoot[C]{\fontsize{8}{10} \selectfont Page \thepage ~of {\hypersetup{hidelinks}\pageref{LastPage}}}
    \fancyhead[L,C,R]{} 
    \setlength{\headheight}{10.0pt}
    \ifnum\AAM=1 
        \fancyhead[C]{\fontsize{8}{10} \selectfont Accepted author manuscript: 
        Barter et al. (2026), 18th International Conference on Agents and Artificial Intelligence (ICAART)}
    \else
        \renewcommand{\headrulewidth}{0pt} 
    \fi
\fi

\section{\uppercase{Introduction}}
\label{sec:introduction}
Bond markets are fundamental to global finance, yet remain comparatively under-explored in artificial intelligence research. Sentiment analysis quantifies the tone or emotional score of textual data, converting qualitative news, reports, or social-media posts into numerical scores. It has long been known that media sentiment can predict market trading volumes \citep{tetlock-2007}, and recent works demonstrate that sentiment analysis can predict bond market returns \citep{bartov2023socialmedia, isakin-2023, pineiro2021influence}. 

Several transformer-based models, such as FinBERT \citep{araci} and FinGPT \citep{wang2023fingpt_benchmark}, have been fine-tuned on financial datasets. Yet, while it is known that bond markets react differently to equity markets on news \citep{isakin-2023}, there are no models fine-tuned specifically for bond market sentiment. To address this gap, we introduce BondBERT, a transformer-based sentiment model fine-tuned on 17,000 bond-related news articles covering 2018--2022. We aim to show how domain-adaptive fine-tuning can extend transformers into domains with inverse sentiment-price relationships, small effect sizes, and noisy signals. 

We benchmark BondBERT against FinBERT, FinGPT, and Instruct-FinGPT across four tasks: (i) correlation with bond returns on sentiment shock days, (ii) event-study directional accuracy, (iii) graphical representation between sentiment and returns, and (iv) bond price prediction using an LSTM model. 

Thus, we address two questions: (i) Does a bond-specific transformer (BondBERT) produce sentiment more aligned with bond price movements than other baselines? and (ii) Does BondBERT sentiment carry predictive power for future bond prices? Results show that BondBERT performs better than the benchmarks, with stronger predictive power, lower normalised RMSE, and higher information coefficient. 

This work introduces the first bond-specific sentiment model, positioning it as a perception and reasoning module for financial decision-support agents. Furthermore, it provides a systematic evaluation of how language models can be adapted to domains with challenging characteristics and where reliable training labels are difficult to obtain. The model's signals can be embedded into forecasting and trading components of decision-support systems, with applications in trading, risk management, and policy analysis.

The remainder of this paper is organised as follows: Section~\ref{sec:lit_review} reviews previous research related to financial sentiment analysis, Section~\ref{sec:exp_setup} describes our dataset and experimental setup, and Section~\ref{sec:methodology} covers the architecture of BondBERT and training procedure. Finally, Sections~\ref{sec:results} and \ref{sec:future_work} present our results and outline future research directions, respectively. 

\section{Literature review} \label{sec:lit_review}
Early approaches to financial sentiment analysis relied on lexicon-based methods (e.g., Loughran-McDonald), which count domain-specific positive and negative words to generate sentiment scores. Although easy to implement, these methods often fail to capture context, negation, and subtle linguistic signals \citep{du2024financial}. Machine learning classifiers such as SVMs and logistic regression improved upon this by learning feature weights from labelled financial text, sometimes using lexicon outputs as inputs. They are generally more accurate than pure lexicons, but these models still depend on manual annotation and often struggle with complex phrasing \citep{peng2024tone}. On the other hand, deep learning models such as RNNs, LSTMs, and CNNs have further advanced the field by automatically learning latent representations that encode word order and context, yielding higher accuracy, but requiring larger datasets and high compute \citep{du2024financial}.

More recently, transformer-based models such as FinBERT have emerged as a benchmark for financial sentiment analysis. Introduced by \citet{araci}, FinBERT extends BERT through additional domain-adaptive pretraining on financial corpora including Reuters, Yahoo Finance, corporate reports, and earnings transcripts \citep{du2024financial}. Comparative studies show that FinBERT outperforms both the base BERT model and traditional machine learning approaches, largely due to its contextual embeddings and financial domain adaptation.\footnote{\url{https://dataloop.ai/library/model/prosusai_finbert/}} Araci's original work demonstrated significant accuracy gains even with limited fine-tuning data \citep{araci}, and subsequent studies confirm its robustness and better performance relative to other large language models across various financial sentiment datasets \citep{fincomms, kang-2025, du2024financial, Huang-2023-FinBERT}.

\begin{figure*}[t]
 \centering
 \includegraphics[width=1\linewidth]{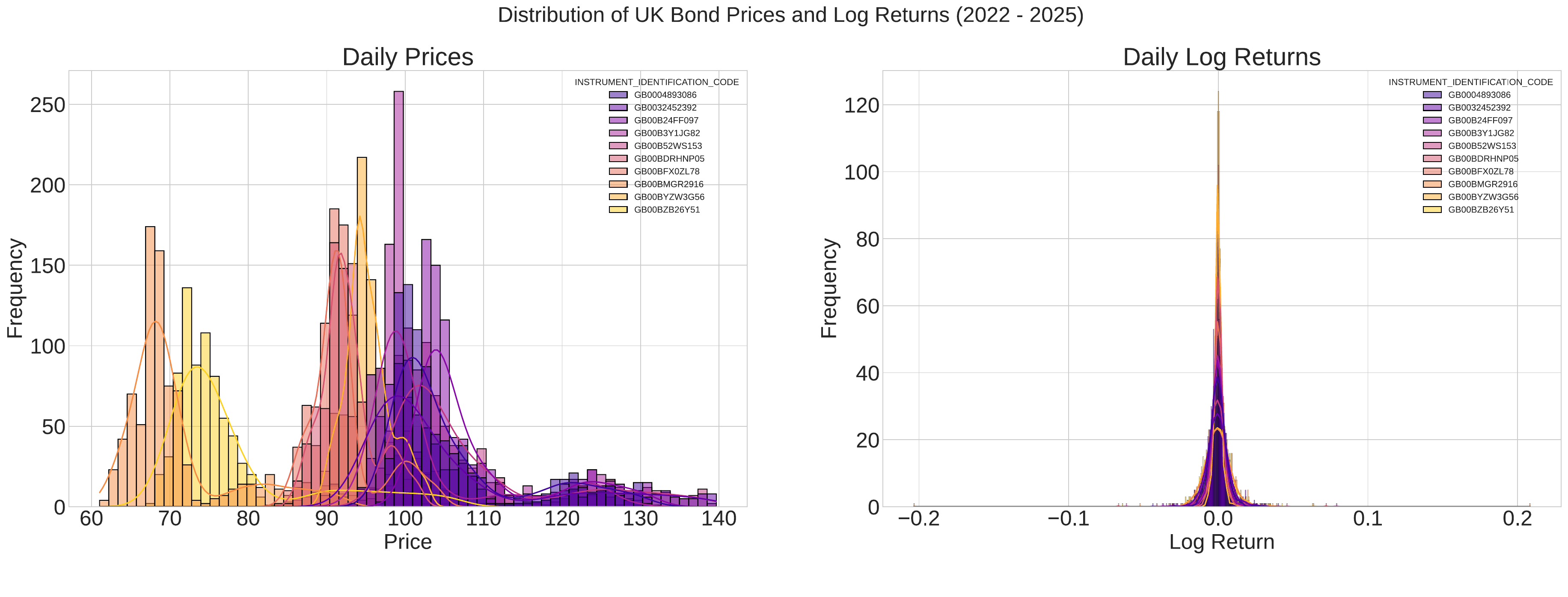}
 \caption{Distribution of price and log returns for the 10 UK instruments.}
\label{fig:bond_distributions}
\end{figure*}

While FinBERT has been widely applied to equity market forecasting \citep{fan-2024, Gu2024Predicting, Cicekyurt-2025, zhang2025dynamic}, its use in bond markets has been limited \citep{makara2024news, liu2025-multilevelsentiment}. Importantly, no variant of FinBERT has yet been trained or adapted specifically for bond-related text. In this work, we address this gap by employing the FinBERT model introduced by \citet{fincomms}, and adapt it to the analysis of sentiment of bond markets (see details in Section~\ref{sec:methodology}). 

The development of FinGPT represents one of the first systematic efforts to create an open-source financial large language model. The framework introduced an automated pipeline for large-scale financial data collection, lightweight adaptation methods such as LoRA/QLoRA, and reinforcement learning with stock prices for alignment \citep{liu2023fingpt_core}. Its four-layer architecture provides a foundation for customisable financial AI research \citep{yang2023fingpt_open}. Building on the above, an instruction-tuning benchmark for FinGPT was proposed, showing that relatively small models (7 billion parameters) fine-tuned with finance-specific instructions can match or even outperform larger untuned models \citep{wang2023fingpt_benchmark}. Thus, recent studies used FinGPT for predictive modelling. \citet{liang2024dissemination} developed a dissemination-aware FinGPT for stock prediction, where company news is grouped to capture dissemination breadth, while context is added to the prompts, achieving an improvement of 8\% in the prediction of short-term movement over LLM using raw news. Moreover, \citet{zhou2025end2end} proposed an end-to-end trading framework that combines FinGPT sentiment pipelines, illustrating the feasibility of incorporating LLMs into trading workflows.

Another popular tool, Instruct-FinGPT, applies instruction tuning to adapt general-purpose LLMs in finance by converting small supervised sentiment datasets into instruction-style data and fine-tuning LLaMA-7B \citep{zhang2023instruct}. Their models outperformed FinBERT and zero-shot ChatGPT in sentiment classification, showing improved performance in context-sensitive inputs. \citet{wang2023fingpt_benchmark} further supported this, confirming the need for instruction-tuned models in finance. Moreover, based on systematic evaluations of instruction-tuned LLMs across finance it was found that instruction tuning consistently improves performance, enabling smaller models to close the gap with larger ones \citep{fatemi2025comparative}. However, limitations remain beyond sentiment analysis, such as long-form document parsing (e.g., annual reports and 10-Ks) and valuation modelling \citep{li2023survey, jadhav-2025}. 

Our work connects to research on decision-support systems, since BondBERT can be viewed as a perceptual agent module embedded within multi-agent financial architectures, where agents handle forecasting, risk assessment, and trading. 

\section{Experimental setup} \label{sec:exp_setup}
We use two primary data sources, bond market transactional data and financial news articles. The transactional bond market data was provided by Propellant.digital, a Software as a Service (SaaS) company that has developed a platform that aggregates and standardises trading data across markets, serving as a working prototype for a consolidated tape in the bond market.\footnote{\url{https://propellant.digital/}} The data was extracted from their Big Query client.api and the subset we used covers all trading days from January 2022 to January 2025. To ensure representative market movements, bonds were filtered for sufficient liquidity and volatility, and daily prices were transformed into logarithmic returns for better analysis. The final dataset includes price information for the 10 most liquid UK sovereign bonds in that period. UK corporate bonds were deliberately excluded from the analysis due to their relative illiquidity.

All bonds are issued with a notational face value (also known as ``par value'' or ``principal''), which is the amount the issuer promises to pay to the bondholder at maturity. Throughout this work, we use the term ``price'' to represent trade price as a percentage of the notational value; for example, if a bond with face value £1000 trades at £1010, we record this trade as (percentage of par) price 101. Price 100 therefore indicates ``trading at par''; prices above 100 indicate ``trading at a premium''; and prices below 100 indicate ``trading at a discount''. Prices tend not to deviate far from the par value 100, and we see this in the distribution of bond trade prices for the 10 instruments shown in Figure~\ref{fig:bond_distributions}. Daily price changes also tend to be small, with most log returns close to zero, as shown in Figure~\ref{fig:bond_distributions}. We have chosen 10 ``liquid'' bonds that are traded frequently, therefore prices react quickly to financial news. In total, the 10 bonds were traded around 30,000 times in January 2022, and trade frequency increased over the period, reaching around 100,000 monthly trades in January 2025. 

In total, nearly 45,000 UK bond-related news articles were collected from 2018--2025, sourced from multiple financial outlets using SerpAPI.\footnote{\url{https://serpapi.com}} Articles were cleaned by removing duplicates, irrelevant topics (e.g., entertainment or sports), and entries with fewer than 500 characters to ensure sufficient length for the articles. Publication dates were standardised to align with the format of the trading data. After cleaning, approximately 30,000 high quality articles remained. From there, we allocated 15,000 articles to training, 3,000 to validation, and 3,000 to testing from 2018 to the end of 2021. We reserve the most recent data (January 2022 to January 2025) for evaluation with the actual bond market prices for that period. This split mimics real world deployment, where models are trained on historical data and tested on future unseen events, avoiding leakage that would influence the performance. 

For reproducibility, we have open-source released all preprocessing scripts, labelling prompts, and model fine-tuning configurations on GitHub.\footnote{\url{https://github.com/Tjbarter/Sentiment-Bond-Analysis}}

\section{Methodology} \label{sec:methodology}
To address the absence of a bond-specific sentiment tool, we developed BondBERT by further adapting a financial domain language model for the fixed-income context. Specifically, we start from the FinBERT model introduced by \citet{fincomms}, which is domain pretrained on a large scale financial text (e.g., SEC filings and Reuters news) but not fine-tuned on labeled sentiment datasets. We fine-tuned this model on our bond-specific sentiment dataset to create BondBERT. For comparison, we use ProsusAI/FinBERT as a baseline,\footnote{\url{https://huggingface.co/ProsusAI/finbert}} which is widely used in prior work and is fine-tuned on the Financial PhraseBank dataset \citep{malo-2014} (general financial news sentiment) but not specialised for bonds. 

The training corpus for BondBERT comprised approximately 17,000 international bond-related news articles spanning 2018--2021, covering macroeconomic, fiscal, and policy domains relevant to bond markets.\footnote{Topics include: ``macroeconomic trends'', ``macroeconomic outlook'', ``bond market'', ``global economic growth'', ``economic outlook'', ``economic uncertainty'', ``interest rates'', ``monetary policy'', ``global risk sentiment'', ``financial market volatility'', ``European bond market'', ``energy crisis'', ``global inflation'', ``global GDP slowdown'', ``macroeconomic report'', ``bond market'', ``investor sentiment'', ``sanctions and trade wars'', and ``geopolitical tensions''.} These articles were cleaned for duplicates, irrelevant topics, and insufficient content length, then aligned to market dates. Sentiment labels were generated using GPT-4.1-nano via the OpenAI API and evaluating different prompts designed to capture the expected impact of each article on bond prices.\footnote{We evaluated different GPT models, such as GPT-4.1 and GPT-4.1-mini, but due to similar results, we opted for GPT-4.1-nano, which is more computationally efficient.}  
 
We used GPT-4.1-nano to efficiently generate sentiment labels, testing prompt formats until we identified one that produced consistent and financially relevant outputs. This was critical because general-purpose prompts often misinterpret fixed income market signals (e.g., interpreting ``interest rate rise'' as positive, even though it is negative for bonds and will lead to bond prices falling). However, we recognise that this approach introduces potential label noise and bias due to the model's prior training data and prompt sensitivity. To mitigate this, we conducted prompt engineering experiments and cross-checked a subset of labels against known market reactions, though future work should incorporate human annotated or hybrid validation to improve model accuracy.

We constructed a set of twelve representative financial news snippets, four clearly negative, four clearly positive, and four neutral, based on real bond market reporting. These examples included diverse contexts such as central bank rate hikes, inflation reports, bond issuance announcements, and fiscal policy updates. Each candidate prompt was evaluated on the basis of its alignment with expected bond price reactions and its consistency across similar examples. 

The best performing prompt was concise and framed the model as a bond market analyst, improving label consistency and ensuring that positive scores reflected bond price increases rather than general economic optimism. We then asked the model to assign a sentiment score between $-1$ (strongly negative for bond prices) and $+1$ (strongly positive), with $0$ indicating neutral. The final prompt, which was deliberately kept concise for better accuracy, was:

\begin{quote}
{\em ``You are a financial analyst. Return only one float between -1.00 and +1.00, indicating the immediate impact on the bond market prices. Negative means prices will fall; Positive means prices will rise.''}
\end{quote}

\noindent
Using this prompt, we processed 17,000 articles for training, almost 2,000 for validation, and 2,000 for testing, in order to generate continuous sentiment scores in the range $[-1,1]$. These scores served as training labels for BondBERT's supervised fine-tuning stage. We merged sentiment scores with the article text, using titles as unique keys to ensure accurate label alignment.

We initially considered further fine-tuning FinBERT on bond-specific sentiment data. However, because FinBERT's prior training was optimised for general financial sentiment, we expected additional fine-tuning might conflict with existing scores and introduce noise rather than improve alignment with bond price movements. For this reason, we developed BondBERT as a separate model, pre-trained on financial text but fine-tuned on bond market data. Sentiment scores were categorised into three categories, negative, neutral, and positive, and stratified undersampling was applied to mitigate class imbalance by creating equal sample sizes across the categories. 

We performed hyperparameter optimisation using Optuna to identify the best configuration for BondBERT. The search explored learning rate, batch size, dropout rates, and activation functions, with the objective of maximising validation accuracy. The final selected parameters, summarised in Table~\ref{tab:bondBERT_parameters}, were used for all subsequent experiments.

\begin{table}[t]
\centering
\caption{BondBERT configuration parameters.}
\label{tab:bondBERT_parameters}
\resizebox{0.35 \textwidth}{!}{%
\begin{tabular}{ll}
\toprule
\textbf{Parameter} & \textbf{Value} \\
\midrule
Epochs & 4 \\
Batch size & 8\\
Learning rate & 2.3e-05\\
Attention probability dropout & 0.1 \\
Hidden activation & gelu \\
Hidden dropout probability & 0.1 \\
\bottomrule
\end{tabular}
}
\end{table}

Text preprocessing was handled using Hugging Face's AutoTokenizer, with a 512-token input limit. Articles exceeding this limit were segmented into multiple chunks, each independently tokenised and scored; final sentiment predictions were derived by averaging the chunk outputs.

Both BondBERT and FinBERT output probabilities for three sentiment classes: negative, neutral, and positive. Following a process similar to \citet{borovkova2022news}, we convert sentiment probabilities into a single continuous sentiment score \(s \in [-1,1]\) using the Normalised Difference Index (NDI):
\begin{equation}
s = \frac{p(\text{positive}) - p(\text{negative})}
{p(\text{positive}) + p(\text{negative})}, 
\end{equation}
\noindent
where $s$ is the sentiment score, p(\text{positive}) is the probability that the text is positive and p(\text{negative}) is the probability that the text is negative. This transformation is particularly useful in our context because it explicitly removes the neutral class and amplifies the polarity of the sentiment signal. By focusing only on positive and negative components, NDI more directly captures information that can move the market. 

To assign sentiment using FinGPT, we obtained the open-source model weights from Hugging Face,\footnote{\url{https://huggingface.co/FinGPT}} which requires specific use licenses. The model was deployed via the official inference endpoint, running the LLaMA-2–7B base model (llama-2-7b-hf-owd) on a NVIDIA L40S GPU (48GB memory). To specialise the backbone for financial applications, we attached the LoRA adapter, FinGPT/fingpt-mt\_llama2-7b\_lora, which provides multi-task instruction tuning optimised for sentiment analysis. 

We used Instruct-FinGPT in its 13B parameter version, accessed through the Hugging Face inference endpoint.\footnote{\url{https://huggingface.co/FinGPT/Instruct-FinGPT-13B}} No LoRA adapters were attached since the model is already instruction-tuned on financial sentiment. Instruct-FinGPT expects input formatted as task-specific instructions followed by a text passage, otherwise outputs are not more reliable than zero-shot prompting. Therefore, for a fair comparison, we used the same BondBERT prompt broken into smaller sentences with guidelines to avoid chattiness: 

\begin{quote}
{\em ``You are a financial analyst. Return only one float in [-1.00, +1.00] indicating the immediate impact on government bond prices. Positive means prices rise (yields fall). Negative means prices fall (yields rise). Do not infer from equities/FX unless explicitly linked to yields/inflation/rates. Output must be a bare number (e.g., 0.2), no words or labels.''}
\end{quote}

\begin{figure*}[t] 
  \centering
  \begin{subfigure}{0.49\textwidth}
    \centering
    \includegraphics[width=\linewidth]{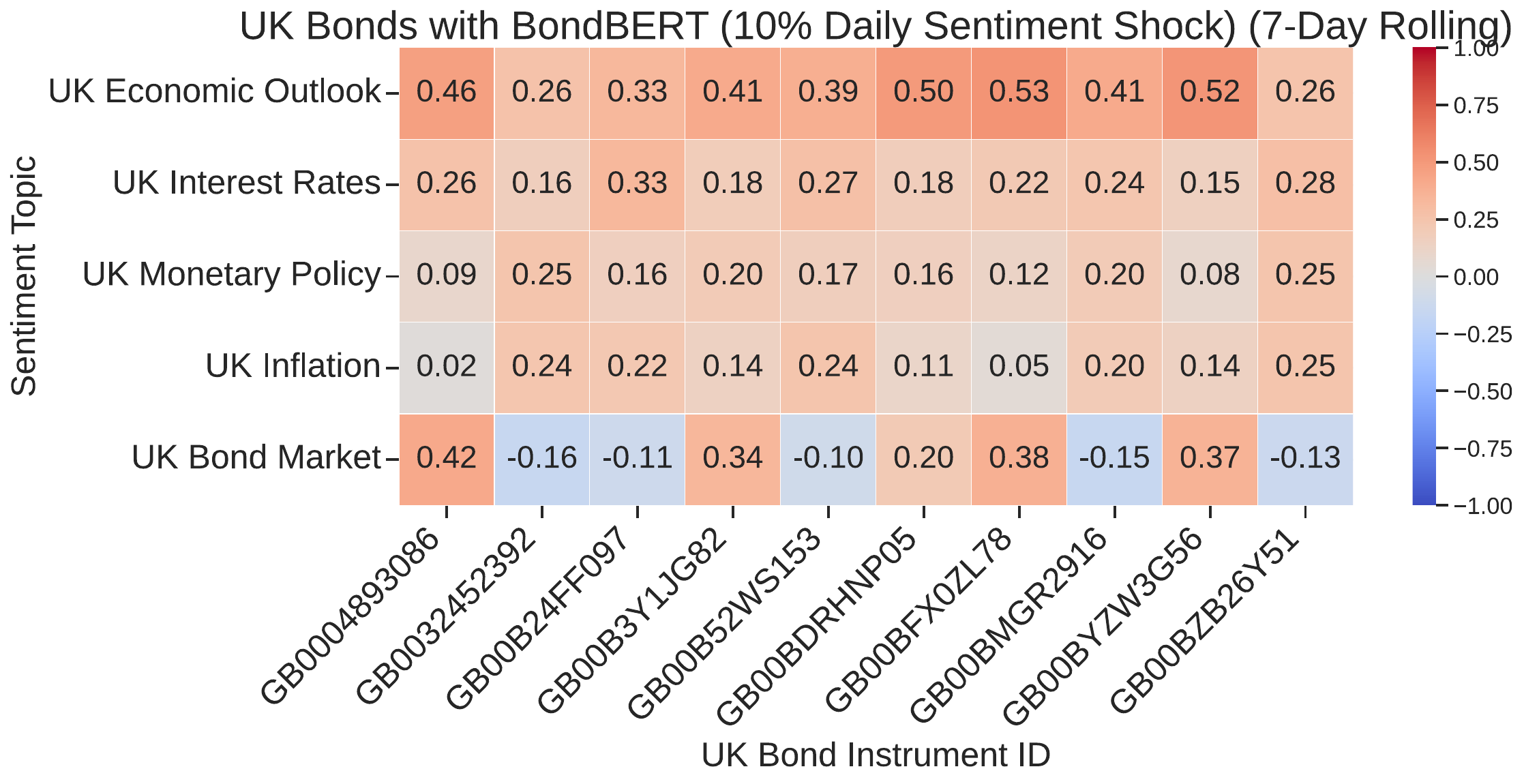}
    \caption{BondBERT}
    \label{fig:uk_only_bondbert_7_day_correlations}
  \end{subfigure}
  \begin{subfigure}{0.49\textwidth}
    \centering
    \includegraphics[width=\linewidth]{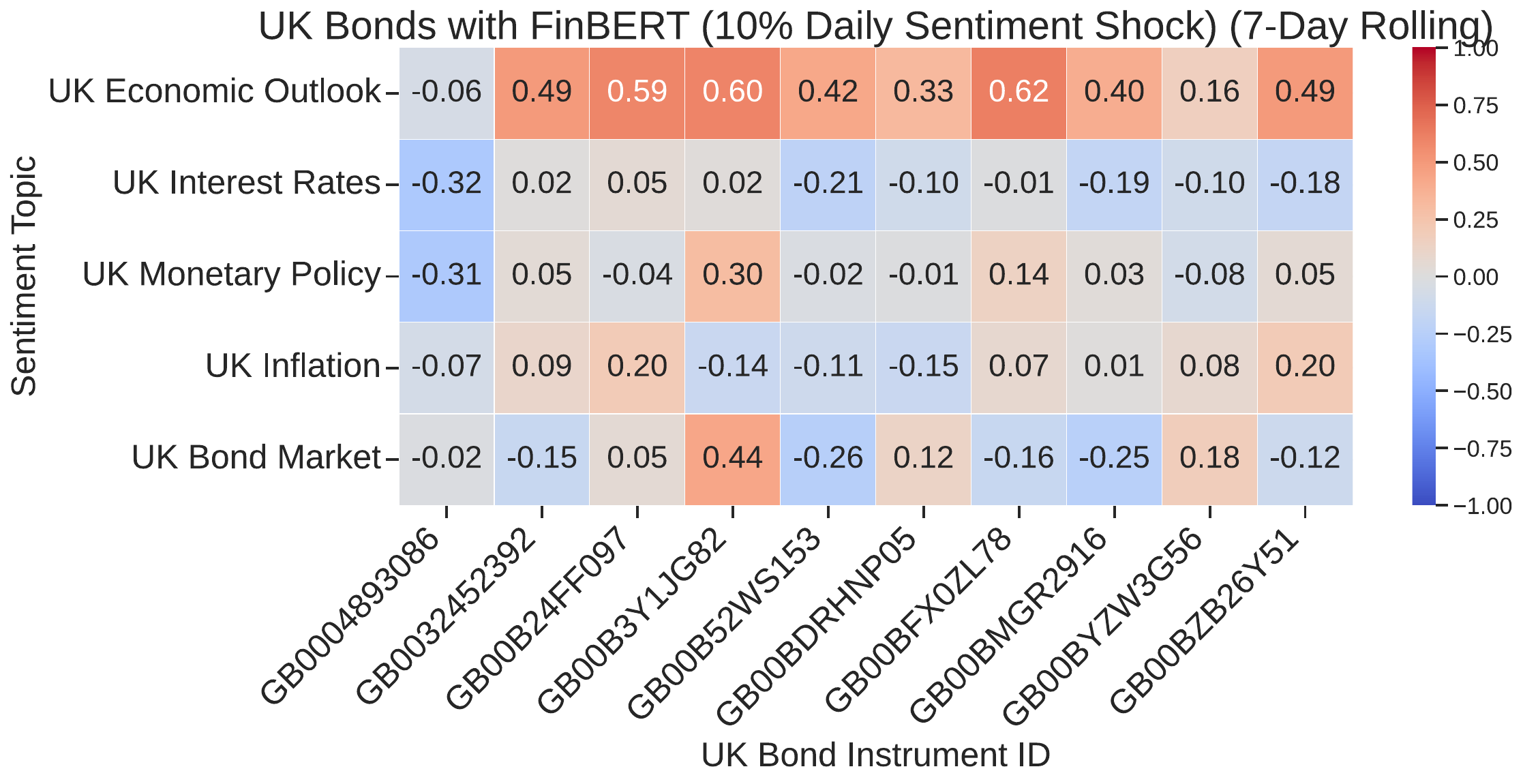}
    \caption{FinBERT}
    \label{fig:uk_only_finbert_7_day_correlations}
  \end{subfigure}\hfill
  \par\medskip
  \begin{subfigure}{0.49\textwidth}
    \centering
    \includegraphics[width=\linewidth]{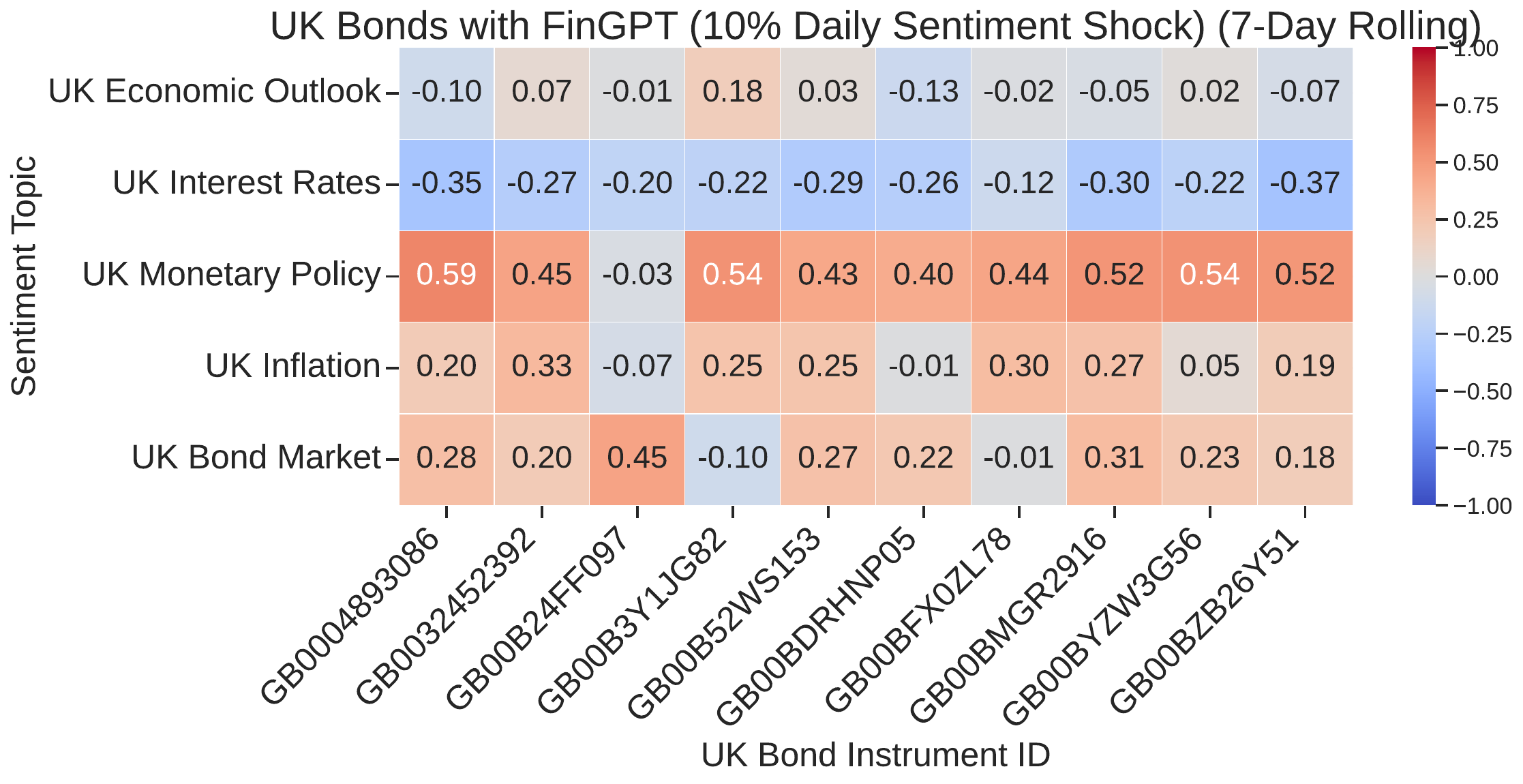}
    \caption{FinGPT}
    \label{fig:uk_only_fingpt_7_day_correlations}
  \end{subfigure}\hfill
  \begin{subfigure}{0.49\textwidth}
    \centering
    \includegraphics[width=\linewidth]{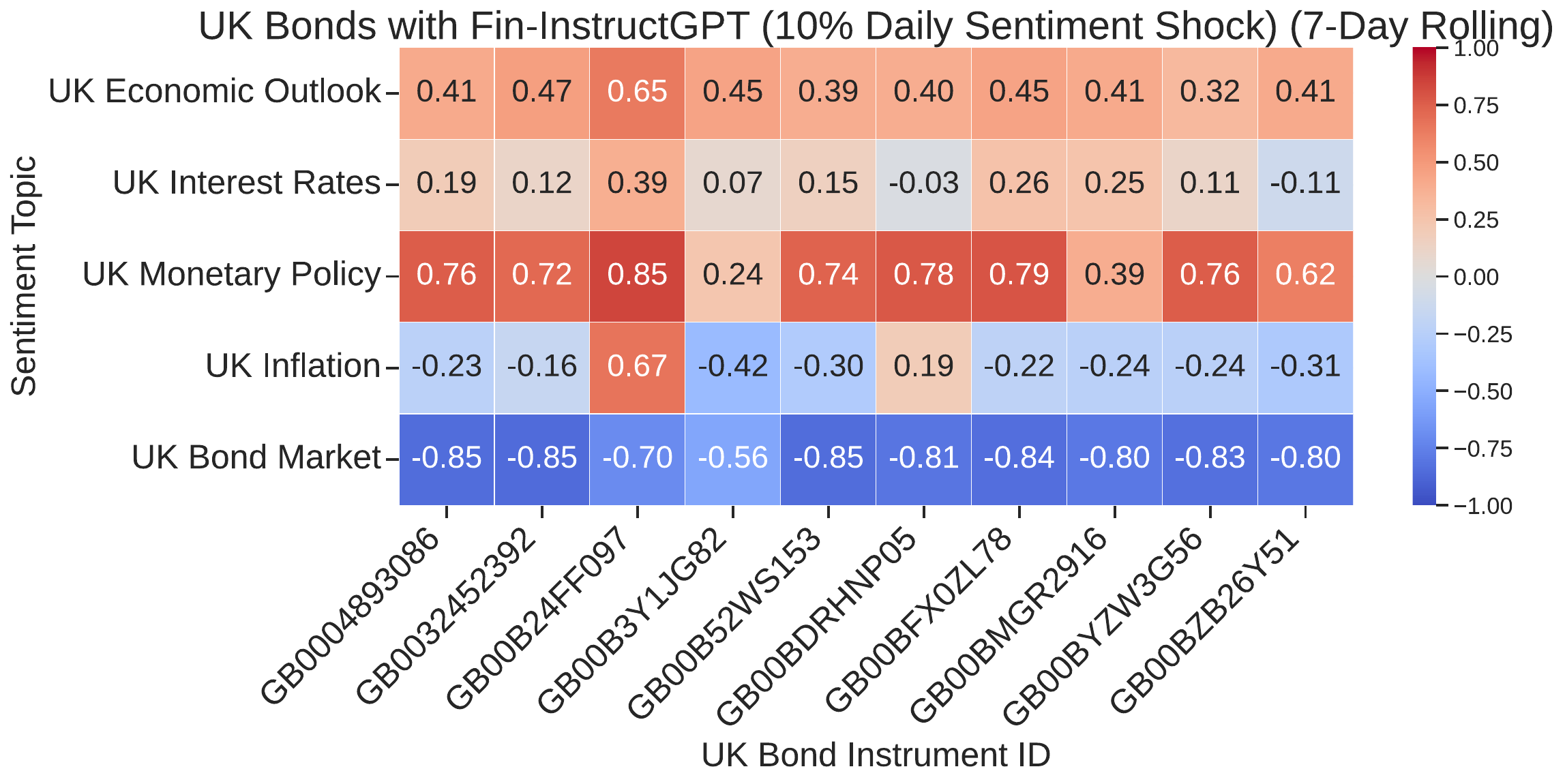}
    \caption{Instruct-FinGPT}
    \label{fig:uk_only_instruct_7_day_correlations}
  \end{subfigure}
  \caption{Sentiment correlations with UK bond prices over a 7-day rolling window across four models.}
  \label{fig:correlations}
\end{figure*}

Finally, it is worth clarifying that because FinGPT and Instruct-FinGPT are not stand-alone architectures, we had to use LLaMA-2 as their underlying backbone. These models were released as adaptations of the LLaMA-2 family (7B and 13B), with LoRA adapters provided for FinGPT and instruction-tuned checkpoints for Instruct-FinGPT specifically designed for this backbone. Although newer open-source backbones such as LLaMA-3 and LLaMA-4 have been released, the FinGPT resources currently available are not compatible with them. Since our goal was to compare BondBERT with other financial models, we included FinGPT and Instruct-FinGPT in the evaluation despite their dependence on LLaMA-2.

One drawback of using GPT-based annotation is the potential risk of look-ahead bias. Since GPT models are trained on large post-event corpora, they may inadvertently leverage knowledge of subsequent market outcomes when labelling historical articles. In our current setup, we did not impose strict temporal framing in the prompt, so we cannot fully rule out this risk. However, the prompt was designed to request the immediate expected impact of each article on bond prices, which partly reduces the chance of future knowledge being incorporated. Future work should address this more rigorously by enforcing time-consistent prompting or complementing synthetic labels with human annotation. This applies not only to GPT-4.1-nano, but also to Instruct-FinGPT and any LLM-based annotator. 

\section{Results} \label{sec:results}
For evaluation, we adopted an event-based methodology to isolate impactful news. Sentiment shocks were defined as daily sentiment values in the top or bottom 10th percentile. When multiple shocks occurred on the same day, their scores were averaged. 

Figure~\ref{fig:correlations} presents Pearson correlations between sentiment scores and UK bond returns on six economic topics, using a 10\% daily sentiment shock (i.e., 10\% most negative and 10\% most positive scores) and a 7-day rolling window. Overall, we see that BondBERT has positive correlations across most topics---with negative correlations only for topic ``UK Bond Market''. FinBERT only has positive correlations for the topic ``UK Economic Outlook''---the other topics are neutral or negative---consistent with the general financial data on which it is trained. FinGPT exhibits positive and negative correlations, which are relatively strong and more consistent than FinBERT. Finally, Instruct-FinGPT displays strong negative correlations for topic ``UK Bond Market'' and weaker, less consistent correlations elsewhere. In general, BondBERT shows the best alignment between bond sentiment and bond price movements, which provides preliminary evidence of the suitability of BondBERT for sentiment analysis of bond markets. 

\begin{figure*}[t]
    \centering
    \includegraphics[width=0.62\linewidth]{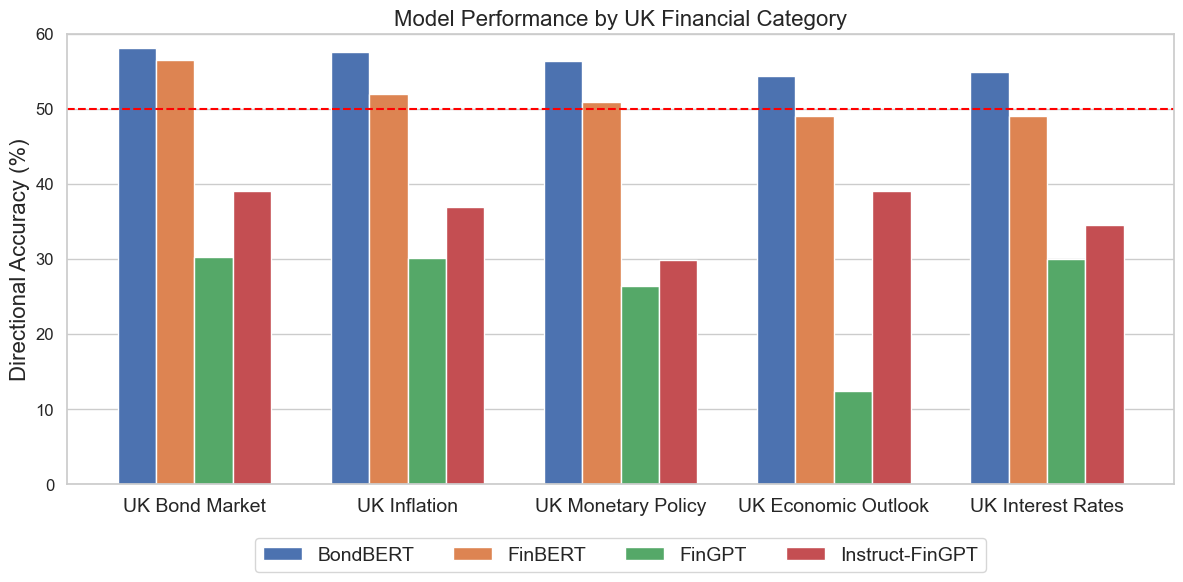}
    \caption{Model performance comparison, showing directional accuracy of sentiment relative to bond returns.}
    \label{fig:model_performance}
\end{figure*}

Figure~\ref{fig:model_performance} compares the directional accuracy of the four models, relative to a 50\% random baseline (horizontal red line). Directional accuracy is measured as the directional prediction (up/down) of the next-day bond returns after a 10\% daily sentiment shock. We see that BondBERT (blue bars) has the highest directional accuracy on all five topics and is the only model that consistently outperforms the random baseline. FinBERT (orange bars) performs second best, but is inconsistent and does not outperform the random baseline on two of the five topics. FinGPT (green bars) and Instruct-FinGPT (red bars) both perform poorly, well below the random baseline on every topic. These results clearly show that BondBERT is the best performing model, and the only model with directional accuracy better than random.

To assess the predictive power of the sentiment models, we used an LSTM to predict next-day bond prices using the fixed-length rolling histories of daily bond transaction prices and the daily sentiment scores from each model as input. All input features and targets were standardised to zero mean and unit variance. The LSTM was optimised using the Adam optimiser, with weight decay and mean squared error (MSE) loss, applying early stopping based on validation loss to avoid overfitting. Hyperparameters such as hidden layer size, number of recurrent layers, dropout, learning rate, weight decay, and history length were optimised using Optuna's Bayesian sampler. Model evaluation is conducted on the unseen test set using normalised RMSE and the information coefficient (IC), and Diebold–Mariano (DM) tests with Newey-West corrected errors to assess whether forecast accuracy differences are statistically significant.

Table~\ref{tab:lstm_results} presents the LSTM results, showing good predictive performance in all models, with low RMSE and moderate to high IC values. BondBERT achieves the lowest normalised RMSE and the highest IC, which suggest that it can be used to predict price movements more accurately than the other models. Table~\ref{tab:dm_results} presents the p-values of the DM test on normalised RMSE values of BondBERT against the other three models.  The null hypothesis of the DM test is that competing forecasts between BondBERT and the competing model have the same population accuracy. We see that the null is rejected in favour of BondBERT in half of the bonds (values in bold), which demonstrates that BondBERT generates statistically significant improvements in normalised RMSE.

In summary, we have shown that BondBERT is able to achieve consistently positive sentiment correlations, which reverses the negative bias exhibited by FinBERT, FinGPT, and Instruct-FinGPT. We have also shown that BondBERT has higher directional accuracy and significantly outperforms other models on next-day price prediction. 

\begin{table}[t]
\centering
\caption{Mean LSTM results per model, showing Normalised RMSE and Information Coefficient (IC).}
\label{tab:lstm_results}
\resizebox{0.47\textwidth}{!}{
\begin{tabular}{lcccc}
\toprule
\textbf{Metric} & \textbf{BondBERT} & \textbf{FinBERT} & \textbf{FinGPT} & \textbf{Inst-FinGPT}\\
\midrule
RMSE & {\bf 0.0079} & 0.0086 & 0.009 & 0.01 \\
IC &{\bf 0.8000} & 0.7450 & 0.747 & 0.67 \\
\bottomrule
\end{tabular}
}
\end{table}

\begin{table}[t]
\centering
\caption{DM test p-values for BondBERT against the baselines. Values in bold denote significance at the 5\% level.}
\label{tab:dm_results}
\resizebox{0.47\textwidth}{!}{%
\begin{tabular}{lccc}
\toprule
\textbf{Instrument ID} & \textbf{FinBERT} & \textbf{FinGPT} & \textbf{Inst-FinGPT}\\
\midrule
GB0004893086 & 0.085 & 0.397 & \textbf{0.014} \\
GB0032452392 & 0.977 & 0.972 & 0.261 \\
GB00B24FF097 & 0.3959 & \textbf{0.04} & \textbf{0.0011} \\
GB00B3Y1JG82 & \textbf{0.0009} & \textbf{2e{-5}} & \textbf{0.01} \\
GB00B52WS153 & 0.1272 & \textbf{0.035} & 0.082 \\
GB00BDRHNP05 & \textbf{0.0042} & \textbf{1e{-14}} & \textbf{1e{-14}} \\
GB00BFX0ZL78 & 0.99 & 0.96 & 0.95 \\
GB00BMGR2916 & \textbf{0.0028} & \textbf{0.0007} & 0.145 \\
GB00BYZW3G56 & 0.2026 & \textbf{3e{-8}} & \textbf{6e{-8}} \\
GB00BZB26Y51 & \textbf{0.0346} & \textbf{0.0153} & 0.2131 \\
\bottomrule
\end{tabular}%
}
\end{table}

\section{Conclusions and Future work} \label{sec:future_work}
This study introduces BondBERT, a bond-specific sentiment model, and demonstrates its advantage over three other tools in aligning sentiment with bond price movements. We showed the importance of a transferable strategy for adapting LLMs to markets where sentiment dynamics diverge from equity assumptions and generalise to other asset classes. BondBERT reverses the wrong-sign bias seen in equity-trained models and delivers more consistent forecasting improvements. The analysis is restricted to UK sovereign bonds, since the illiquidity of corporate bonds prevents a meaningful evaluation, and the model is trained on synthetic labels generated by GPT-4.1-nano. Although this approach is cost-effective and scalable, it may not fully capture domain expertise and constrains label fidelity. We plan to introduce human annotation and semi-supervised refinement to improve label quality. Moreover, our goal is to expand the dataset across, incorporating international sovereign and corporate bonds with higher liquidity. These extensions will test the generalisability of BondBERT and strengthen its potential applications in trading, risk management, and policy. 

Beyond finance, we highlight a broader AI challenge, learning under weak effect sizes and noisy supervision, where sentiment diverges from intuitive human interpretations. While BondBERT is not a standalone agent, it demonstrates how specialised perception modules can be embedded into decision-support agents for fixed income markets and interact with forecasting and trading agents as part of multi-agent systems, extending the role of NLP in financial AI. In the longer term, the goal is to incorporate predictive models into an agentic-AI framework to provide bespoke investment advice to individual investors via a mobile application. For more details, see \ifnum\BLIND=1{[REDACTED FOR BLIND REVIEW].}\else{the artificial intelligence for collective intelligence (AI4CI) vision, described in \cite{ai4ci-position-2024-fullnames}.}\fi


\section*{\uppercase{Acknowledgements}}

\ifnum\BLIND=1
    Redacted for blind review.
\else
    This work was supported by UK Research and Innovation (UKRI) Engineering and Physical Sciences Research Council (EPSRC) [grant number EP/Y028392/1]: AI for Collective Intelligence (AI4CI). Toby Barter was supported by a Faculty of Engineering Summer Research Internship. The authors thank Vincent Grandjean and Oliver Haste at Propellant Digital for expert advice and data access. 
\fi

\bibliographystyle{apalike}
{\small
\bibliography{bond-bert}}

\end{document}